%% file: main.tex
\documentclass[12pt]{iopart}

\usepackage{iopams,setstack} 
\usepackage{harvard} 
\usepackage{graphicx}
\usepackage{subfigure}
\usepackage{booktabs}

\usepackage{color}

\mathchardef\mhyphen="2D

\begin{document}

\note[Validation of Geant4 nuclear models using the dose distribution of a proton pencil beam]{Validation of nuclear models in Geant4 using the dose distribution of a 177~MeV proton pencil beam}

\author{David C Hall$^{1}$\footnote{\label{foot:equal}Both authors have contributed equally to this work.}, Anastasia Makarova$^{2}$\footnotemark[\value{footnote}], Harald Paganetti$^{1}$, Bernard Gottschalk$^{3}$}
\address{$^{1}$Department of Radiation Oncology, Massachusetts General Hospital and Harvard Medical School, Boston, Massachusetts 02114, USA}
\address{$^{2}$NCTeam, Department of Radiation Oncology, Medical Faculty, University of Duisburg-Essen 45122, Germany}
\address{$^{3}$Harvard University Laboratory for Particle Physics and Cosmology, 18 Hammond St., Cambridge, Massachusetts 02138, USA}
\eads{\mailto{dchall@mgh.harvard.edu}, \mailto{anastasia.makarova@uni-due.de}}

\begin{abstract}
A proton pencil beam is associated with a surrounding low-dose envelope, originating from nuclear interactions. It is important for treatment planning systems to accurately model this envelope when performing dose calculations for pencil beam scanning treatments, and Monte Carlo (MC) codes are commonly used for this purpose. This work aims to validate the nuclear models employed by the Geant4 MC code, by comparing the simulated absolute dose distribution to a recent experiment of a 177~MeV proton pencil beam stopping in water.

Striking agreement is observed over five orders of magnitude, with both the shape and normalisation well modelled. The normalisations of two depth dose curves are lower than experiment, though this could be explained by an experimental positioning error. The Geant4 neutron production model is also verified in the distal region. The entrance dose is poorly modelled, suggesting an unaccounted upstream source of low-energy protons. Recommendations are given for a follow-up experiment which could resolve these issues.
\end{abstract}

\submitto{\PMB}

\section{Introduction}
\label{sec:intro}
\input{intro}

\section{Methods}
\label{sec:method}
\input{method}

\section{Results}
\label{sec:results}
\input{results}

\section{Discussion}
\label{sec:disc}
\input{discussion}

\section{Conclusions}
\label{sec:concl}
\input{conclusion}

\section*{Acknowledgements}
The authors thank Dr Juliane Daartz and Dr Ben Clasie for their description of the MGH experimental beam line and Dr Jan Sch\"{u}mann for comments that greatly improved the manuscript.
DH and HP wish to acknowledge support from National Cancer Institute grant U19CA21239. BG thanks the Physics Department of Harvard University for continuing support. AM thanks her PhD supervisor Dr Wolfgang Sauerwein and the Department of Radiation Oncology at University of Duisburg-Essen.

\section*{References}
\bibliographystyle{jphysicsB_withTitles}
\bibliography{biblio}

\end{document}

%% file: intro.tex
Pencil beam scanning (PBS) is a mode of proton therapy whereby each radiation field is constructed from a large number of narrow pencil beams. It is becoming increasingly popular, and in future most proton therapy treatments are likely to be delivered by this treatment technique.

In a PBS radiation field, the absolute dose at a point of interest (POI) in a patient or phantom is computed by adding the contributions of those pencil beams sufficiently near the POI to influence it. In defining this distance of influence, it is important to consider that the primary core of each pencil beam is laterally surrounded by a low-dose envelope. Following the terminology of \citeasnoun{Gottschalk:2015}, the \textit{core} describes the bulk energy deposited by primary protons, the surrounding \textit{halo} is caused by charged secondaries originating from hard nuclear interactions, and the \textit{aura} is the very low dose bath caused by neutral secondaries. In practical beam lines the low-dose envelope may be further complicated by dose due to scattering from upstream apparatus, known as \textit{spray}. Unlike the halo and aura, which are due to basic physics, spray is reducible in principle, though this can prove difficult to achieve in practice. The halo, aura and spray remove energy (integrated dose) from the core. Neglecting this redistribution of dose can lead to absolute dose errors of order 10\% \cite{Pedroni:2005}, though the exact result depends strongly upon field size, spot weighting, energy and depth.

To correctly calculate absolute dose in PBS, as required when commissioning a treatment planning system (TPS), it is necessary to model the halo, and spray if present, with sufficient accuracy. Although the TPS model may be entirely experiment-based \cite{Pedroni:2005}, Monte Carlo (MC) simulations commonly play a large role \cite{Soukup:2005,Parodi:2013}. MC particle transport codes, such as Geant4 \cite{Geant4}, FLUKA \cite{FLUKA} and MCNPX \cite{MCNPX} are regularly used as a gold standard against which TPSs are validated \cite{Clasie:2012,Grevillot:2012,Grassberger:2015}.

It is therefore important to validate the nuclear interaction models employed by these MC codes in the clinical incident energy regime 70--250\,MeV. Recently, a comprehensive absolute measurement of the low-dose region (\citeasnoun{Gottschalk:2015}, hereinafter Go15) explored its complete radial and depth dependence at 177\,MeV. We propose to test the nuclear models employed by Geant4 against this dataset, 295 values of log$_{10}$(D/(MeV/g/p)), whose range of radii, depths and dose variation (five orders of magnitude) affords a test of unprecedented simplicity and scope.

To justify that claim we discuss some competing studies, limiting ourselves to ones whose signal is largely or entirely from nuclear interactions. That excludes studies such as \citeasnoun{Kurosu:2014} which concern percent depth-dose (PDD) distributions in broad beams, where the contribution of nuclear reactions is of order 10\% \cite{Pedroni:2005}.

Experiments with multi-layer Faraday cups \cite{Gottschalk:1999,Paganetti:2003} satisfy our criterion. They are absolute, and the nuclear and EM signals are completely separated. However, they only measure the range distribution of charged secondaries, a 1D test with transverse information integrated out. And they measure the {\em number} of charged secondaries rather than the dose.

More directly comparable to Go15 are measurements of the transverse dose profile which, beyond the core, is attributable either to nuclear secondaries or spray. \citeasnoun{Sawakuchi:2010:exp} measured transverse profiles in air at isocenter and at selected depths in water at 72.5, 148.8 and 221.8\,MeV. They used a small cylindrical ion chamber as well as optically stimulated luminescent detectors and radiochromic film. Their data were later compared with MCNPX using the Bertini cascade model followed by the preequilibrium model \cite{Sawakuchi:2010:mc}. Good agreement of the shape of the MC simulation with experiment was found, over three decades of dose, for the transverse profiles in water at midrange.

More recently \citeasnoun{Lin:2014} measured transverse profiles at six energies between 100 and 225~MeV using radiochromic film at selected depths in a Solidwater phantom. Their IBA dedicated PBS nozzle was not modelled. Rather, the beam at the phantom entrance was approximated by adding to the core two subsidiary Gaussians whose energy dependent widths and amplitudes were fitted to in-air profile measurements. Simulations were performed with TOPAS using the default physics list and two options for the Geant4 EM model. Limiting ourselves to aspects relevant here, one can conclude from their Fig.\,4 and text that TOPAS agrees well with the shape of their measured 2D transverse dose distribution except for the 0.01\% isodose at shallow depths where the double Gaussian approximation to spray is inadequate.

In short, nuclear models in MCNPX and Geant4 have already been tested to the extent that any significant failure to predict the {\em shape} of selected transverse dose profiles would have been detected. They also confirm the expected cylindrical symmetry of the profiles wherever nuclear interactions dominate over spray. Go15, however, provides a test which is both simpler and more comprehensive. Measurements are in water, avoiding issues of the water equivalence of Solidwater. They employ a thimble chamber whose absolute calibration is $^{60}$Co-based as in clinical practice, avoiding issues that arise when using radiochromic film for absolute dosimetry. Measurements span the entire low-dose volume, instead of selected depths. Most important, the absolute dose {\em per incident proton} is reported. These data are therefore uniquely suited to answering the fundamental yet simple question: Does the MC under test correctly predict the dose per incident proton in the region where that dose is dominated by nuclear interactions?

%% file: method.tex
\subsection{The experimental dataset}
\label{sec:method:exp}

The two-dimensional dose distribution of a 177~MeV proton pencil beam stopping in water was measured using the experimental beam line at Massachusetts General Hospital (MGH). The experiment is described in \citeasnoun{Gottschalk:2015} and further detailed in \citeasnoun{Gottschalk:2014:long}.

The experimental setup comprised a proton beam directed at a water phantom. The beam exited a 6.35~cm diameter beam pipe through a 0.008~cm Kapton\circledR{} vacuum window. It then traversed a 150~cm helium bag and a 30~cm air gap before entering the water through a 0.86~cm PMMA tank wall (1.02~cm water-equivalent thickness). An air-filled plane-parallel ionisation chamber containing a total thickness of 0.0064~cm of aluminium was placed on the upstream surface of the tank wall, acting as a beam monitor. An Exradin T1 thimble ionisation chamber measured the dose at various positions within the phantom. The design minimised spray since the beam pipe was significantly larger than the beam width and only homogeneous materials were traversed.

In order to make absolute measurements of dose, the beam monitor was calibrated using a Faraday cup and the thimble chamber was calibrated by an Accredited Dosimetry Calibration Laboratory. Together, these could precisely measure a distribution of dose per proton. By recording depth dose curves at different radial distances from the beam axis (rather than lateral dose profiles at different depths) the signal range was limited. Then the beam fluence could be adjusted at each radial distance, in order to optimise the sensitivity of the experiment. This yielded a dose uncertainty of 3\% and a positional uncertainty of 1~mm.

The model dependent fit of Go15 to their data yielded (among other things) properties of the incident
beam at the tank entrance: beam range $d_{80} = 21.32 \pm 0.13$~cm and beam width $\sigma_x = 5.9 \pm 2.8$~mm.\footnote{Errors taken directly from the fitting program. Even the detailed paper \cite{Gottschalk:2014:long} did not give estimated parameter errors as the authors considered them unreliable in such a complicated fit. Of course, a targeted experiment could more precisely determine the beam parameters.} The angular spread (beam divergence), emittance and width of the Bragg peak were so poorly constrained as to be essentially unmeasured. Using radiochromic film, the beam was found to be cylindrically symmetric $(\sigma_x = \sigma_y)$ to within a few percent.

\subsection{Monte Carlo simulation}
\label{sec:method:mc}

We simulated the two-dimensional dose distribution using Geant4 version 10.1p02 \cite{Geant4}, which was the latest release at the time of writing. A total of $1 \times 10^{8}$ proton histories were simulated.

The simulated geometry was simplified with respect to the experimental setup. The beam source was positioned on the edge of a large volume of water, with the tank wall replaced by its water-equivalent thickness. Upstream apparatus was neglected. The dose distribution was scored in a set of concentric tubes centred on the beam axis, in order to exploit the symmetry of the simulation. The depth and radial bin widths were 1~mm and 4~mm respectively, to match the dimensions of the ionisation chamber. Since the beam source was placed at the surface of the phantom, the absolute dose was simply computed by dividing the scored dose by the number of proton histories simulated.

In the experimental paper, a beam energy of 177~MeV was determined via Janni's range-energy tables \cite{Janni}. However, the mean ionisation energy of water employed by default in Geant4 is taken from revised ICRU-73 tables \cite{ICRU73:revised}, which provide shorter ranges than Janni's tables. Therefore the beam energy was tuned to 177.9~MeV, where the simulated range agreed with experiment. It has been shown in a Monte Carlo study \cite{Peeler:2012} that the dose distribution at large radial distances is unrelated to the size of the core. The width of the Gaussian beam energy distribution is calculated using an experiment-based parametrisation derived specifically for the MGH beam line \cite{Clasie:2012},
\begin{equation}
	\frac{\Delta E}{E} = (-8.6508521 \cdot d_{80}^2 + 51.869336 \cdot d_{80} + 9337.7444) \times 10^{-6}
\end{equation}
where $d_{80}$ is the beam range. This yields a relative energy spread of 0.65\%.

The lateral and angular distributions of the beam were described by Gaussian distributions, with standard deviations $\sigma_{x}$ and $\sigma_{\theta}$ respectively. Although these parameters were estimated by the fitting procedure in the experimental paper, they were poorly constrained. In the simulation, they were tuned to $\sigma_{x} = 5.4$~mm and $\sigma_{\theta} = 18.4$~mrad in order to yield the best description of the experimental core data ($r \leq 1$~cm). Positional and angular distributions were assumed uncorrelated.\footnote{If this assumption were invalid, the model-dependent fit in Go15 would have yielded a statistically significant emittance.}

Appropriate Geant4 physics settings for proton therapy applications have been identified in version 8.1p01 by \citeasnoun{Jarlskog:2008} and in version 9.2 by \citeasnoun{Grevillot:2010}. In this study, similar physics modules were selected in Geant4 version 10.1p02. Electromagnetic interactions were described by the standard model (\verb|G4EmStandardPhysics_option4|), elastic nuclear interactions were described by the standard hadronic elastic physics model (\verb|G4HadronElasticPhysics|), and inelastic nuclear interactions were described by a binary cascade model followed by a pre-compound model of nuclear de-excitation (\verb|G4HadronPhysicsQGSP_BIC_HP|). Additionally, the \verb|G4EmExtraPhysics|, \verb|G4StoppingPhysics|, \verb|G4IonBinaryCascadePhysics| and \verb|G4NeutronTrackingCut| physics modules were used. The range threshold for secondary particle production was set to 0.1~mm.

%% file: results.tex
\begin{figure}
	\centering
	\includegraphics[width=0.7\textwidth]{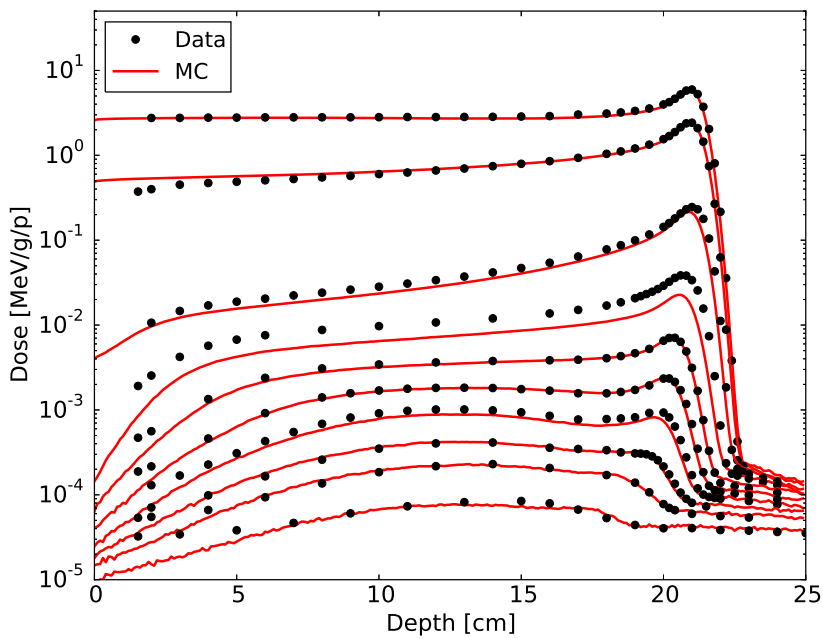}
    \caption{Absolute depth dose curves produced by a 177~MeV proton pencil beam stopping in water, recorded at radial distances of 0, 1, 2, 3, 4, 5, 6, 7, 8 and 10~cm from the central beam axis (in order of decreasing dose). The Geant4 MC simulation (lines) is directly compared to experimental data (circles).}
    \label{fig:depth:log}
\end{figure}
\begin{figure}
	\centering
    \includegraphics[width=0.7\textwidth]{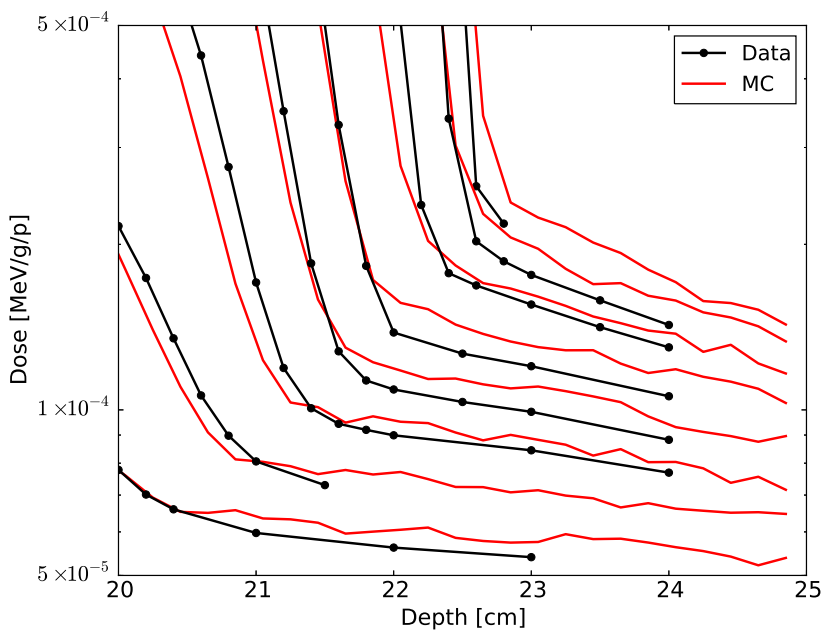}
    \caption{Zoomed view of the distal data of figure~\ref{fig:depth:log}. Data along the beam axis ($r = 0$~cm) are not shown as no experimental measurements were made in this region.}
    \label{fig:depth:distal}
\end{figure}
\begin{figure}
	\centering
    \includegraphics[width=0.7\textwidth]{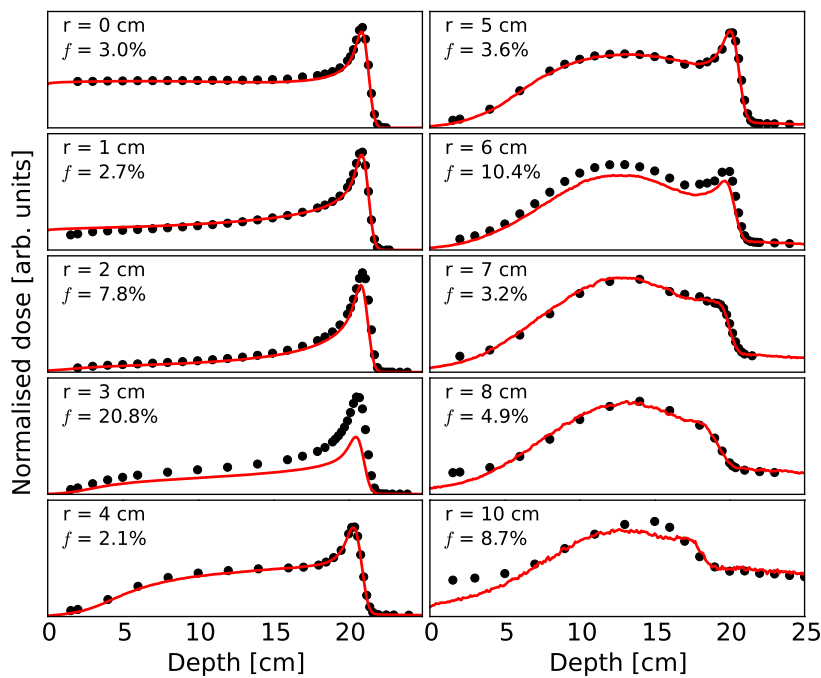}
    \caption{The data of figure~\ref{fig:depth:log} in linear presentation. Each panel is separately normalised to the maximum experimental dose, but the relation between simulation (line) and experiment (circles) is unaltered and absolute. The figure-of-merit $f$ (see text) quantifies the disagreement, expressed in percent.}
    \label{fig:depth:lin}
\end{figure}
\begin{figure}
	\centering
	\includegraphics[width=0.495\textwidth]{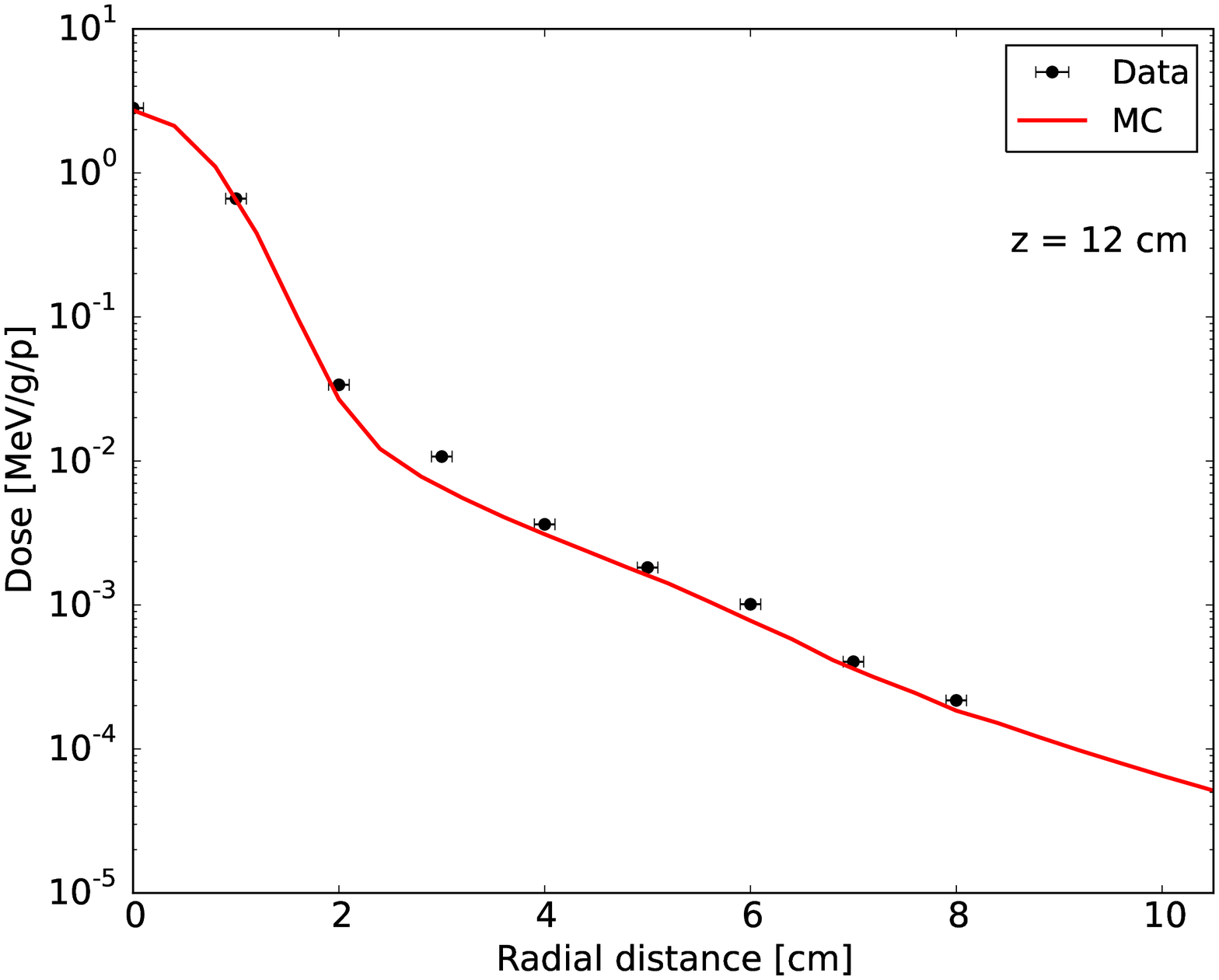}
    \hfill
    \includegraphics[width=0.495\textwidth]{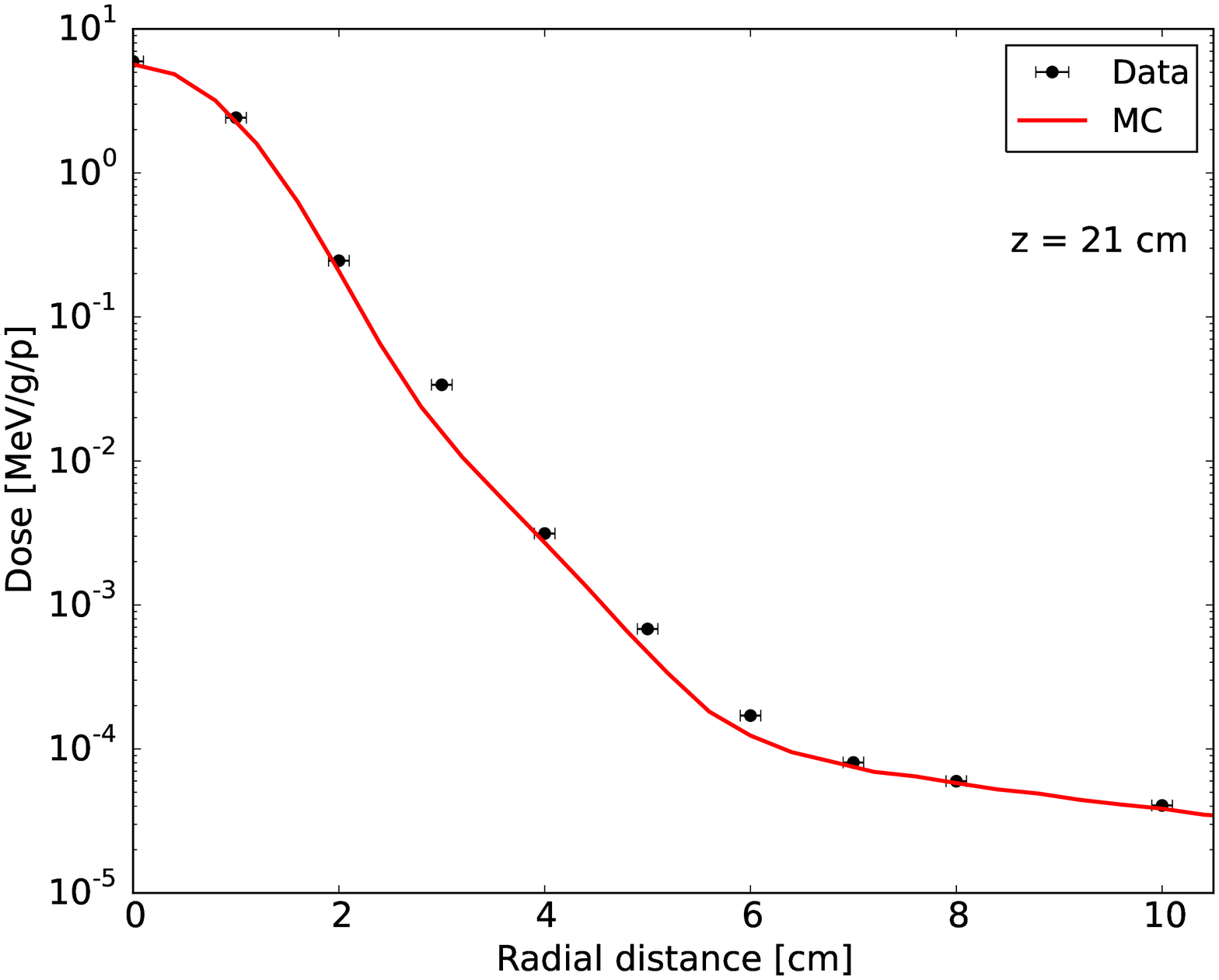}
    \caption{Lateral profiles of the absolute dose distribution produced by a 177~MeV proton pencil beam stopping in water, recorded at a depth of ($a$) 12~cm (mid-range) and ($b$) 21~cm (near end of range). The Geant4 MC simulation (line) is directly compared to experimental data (circles), and radial positioning uncertainties of 1~mm are displayed.}
    \label{fig:radial}
\end{figure}

Figure~\ref{fig:depth:log} presents ten absolute depth dose curves produced by a 177~MeV proton pencil beam stopping in water. These were recorded at radial distances of 0, 1, 2, 3, 4, 5, 6, 7, 8 and 10~cm from the central beam axis, with greater radial distance corresponding to lower dose. Emphasis is given to the distal data in figure~\ref{fig:depth:distal}. The data are also displayed on a linear dose scale in figure~\ref{fig:depth:lin}. Lateral dose profiles at depths of 12~cm (mid-range) and 21~cm (near end of range) are shown in figure~\ref{fig:radial}. In analytical TPSs, these lateral distributions are usually described by a double Gaussian \cite{Yi:2012}.

Disagreement between simulated and experimental depth dose curves is quantified using a figure-of-merit. This is the root-mean-square dose deviation normalised to the maximum measured dose for the depth dose curve under consideration,
\begin{equation}
	f = \frac{\sqrt{\langle (D_{\mathrm{data}} - D_{\mathrm{MC}})^2 \rangle}}{\max (D_{\mathrm{data}})}
\end{equation}
where the calculation includes only depths at which an experimental measurement $D_{\mathrm{data}}$ is available. At each depth, the simulated dose $D_{\mathrm{MC}}$ of the nearest voxel was used. This is displayed in figure~\ref{fig:depth:lin} for each depth dose curve.

%% file: discussion.tex
Before addressing issues evident in the results, we note that overall agreement between Geant4 and experiment is striking, especially considering that \textit{neither dataset is normalised}. By considering the agreement in different regions of the dose distribution, it is possible to verify the modelling of each component: core, halo, aura and spray.

\subsection{Core}
\label{sec:disc:core}

Agreement in the core (radii 0 and 1~cm), which tests the Geant4 EM model, is almost exact. This is unsurprising since the incident beam parameters were optimised to fit the experimental data in this region (enabling the Geant4 nuclear model to be tested at larger radii).

\subsection{Halo}
\label{sec:disc:halo}

The halo involves two distinct physics mechanisms. Coherent scattering (the proton interacts with the nucleus as a whole) gives rise to a Bragg peak, a hint of which is seen as far out as $r = 8$~cm. Incoherent scattering (the proton interacts with constituents of the nucleus) produces the mid-range bump which appears at $r = 4$~cm and persists beyond $r = 10$~cm. Figure~\ref{fig:depth:lin} demonstrates that Geant4 accurately describes both the shape and normalisation of these features, with notable exceptions at $r = 3$ and 6~cm.

\citeasnoun{Gottschalk:2015} already commented on the $r = 6$~cm scan being an outlier, evidence of positioning errors of order $\pm 1$~mm in the experiment. The much larger disagreement at $r = 3$~cm was unexpected and would require a radial positioning error of 3~mm to achieve agreement. On the other hand, it seems unlikely that Geant4 would mismodel these two features, of very different physical origin, by nearly the same factor. This question would be resolved by improved positioning precision and intervening scans out to $r = 5$~cm.

The halo produced by other Geant4 nuclear models was also investigated, though the binary cascade model clearly offered the best description. The Bertini cascade model was found to overestimate incoherent scattering at intermediate radii (increasing the figure-of-merit $f$ by factors up to 3). The Liege intranuclear cascade model was found to underestimate incoherent scattering at large radii (increasing $f$ by factors up to 5). The choice of de-excitation model was not found to have a significant effect.

\subsection{Aura}
\label{sec:disc:aura}

Both experiment and simulation exhibit a clear plateau just downstream of the distal edge of the Bragg peak (see figures~\ref{fig:depth:log} and \ref{fig:depth:distal}). This flat region is attributed to neutral secondaries (neutrons and $\gamma$-rays). Good agreement is observed, though Geant4 is generally higher than experiment by $\sim$10\%. This mismodelling is reduced at $r = 3$~cm, lending further support to the positioning error hypothesis described in section~\ref{sec:disc:halo}. This is a qualitatively new test of MC neutron production.

MC neutron production is, of course, an ingredient in every shielding study, but there it is complicated by neutron transport through complex geometries of various intervening materials. Factor-of-two accuracy is the norm, and is in fact acceptable from a practical viewpoint \cite{Clasie:2010}. In contrast, figure~\ref{fig:depth:distal} is the most direct test possible of the neutron source model in Geant4, with minimal absorbing material, itself water. It shows that Geant4 performs well in both producing and transporting neutrons within water.

\subsection{Spray}
\label{sec:disc:spray}

In contrast, Geant4 is significantly lower than experiment in the proximal region at intermediate and large radii (see figure~\ref{fig:depth:log}). One immediately suspects spray, and this was investigated. We modelled the upstream apparatus described in section~\ref{sec:method:exp}, and moved the MC beam source from the tank wall to the vacuum window. The beam parameters were reoptimised to fit the experimental core dose, but no significant improvement was observed in the proximal region. Either Geant4 is indeed wrong here, or there is an unmodelled source of spray, perhaps scattering of protons in the beampipe originating far upstream. This could be investigated using in-air transverse scans in the entrance region.

%% file: conclusion.tex
We have compared the full dose distribution of a 177 MeV proton beam stopping in water with Geant4 simulations and found absolute agreement over five orders of magnitude, typically within a few percent. The depth dose curves at $r=3$ and 6\,cm are outliers, with Geant4 lower than experiment by as much as 40\%, though even here the shape is well modelled. The discrepancy at $r=3$~cm could be due to experimental positioning error $\approx3$~mm, unlikely but not impossible, or it could mean that G4 underpredicts secondaries at intermediate radii.

Agreement in the aura is also good, marking a new test of MC neutron production. In contrast, measurements at the very lowest entrance doses are substantially higher than Geant4. That suggests a broad unaccounted-for source of low-energy incident protons (spray), although it is conceivable instead that G4 underpredicts back- and transverse scattering.

Our findings call for an improved experiment with better positional accuracy, scans at more intermediate radii, and in-air scans in the entrance region to investigate spray. Eventually, measurements should also be made at other beam energies. That being said, with the exceptions just noted, this study validates the Geant4 nuclear model to a few percent absolute error at a typical proton therapy energy, justifying its use in commissioning PBS TPSs. It also shows that Geant4 is accurate enough to provide kernels modelling the halo for TPSs. It would be interesting to see whether other widely used MC codes, such as MCNPX and FLUKA, are as accurate.